\documentclass[aps,prb,twocolumn,superscriptaddress,showpacs,showkeys]{revtex4-2}

\usepackage{color}
\usepackage{graphicx}% Include figure files
\usepackage{dcolumn}% Align table columns on decimal point
\usepackage{bm}% bold math
\usepackage{float}
\usepackage[mathlines]{lineno}% Enable numbering of text and display math
\usepackage{tabularx}
\usepackage[colorlinks,linkcolor=blue,anchorcolor=blue,citecolor=blue,urlcolor=blue,hyperindex,CJKbookmarks]{hyperref}
\usepackage{footnote}
\usepackage{amsmath}
\usepackage{txfonts}

\begin{document}

\title{Absence of phonon-mediated superconductivity in La$_3$Ni$_2$O$_7$ under pressure}

\author{Zhenfeng Ouyang}\affiliation{Department of Physics and Key Laboratory of Quantum State Construction and Manipulation (Ministry of Education), Renmin University of China, Beijing 100872, China}
\author{Miao Gao}\email{gaomiao@nbu.edu.cn}\affiliation{Department of Physics, School of Physical Science and Technology, Ningbo University, Zhejiang 315211, China}
\author{Zhong-Yi Lu}\email{zlu@ruc.edu.cn}\affiliation{Department of Physics and Key Laboratory of Quantum State Construction and Manipulation (Ministry of Education), Renmin University of China, Beijing 100872, China}

\date{\today}

\begin{abstract}
A recent experimental study announced the emergence of superconductivity in La$_3$Ni$_2$O$_7$ under pressure, with the highest observed superconducting transition temperature ($T_c$) reaching approximately 80 K beyond 14 GPa. While extensive studies have been devoted to the electronic correlations and potential superconducting pairing mechanisms, there lack investigations into the phonon properties and electron phonon coupling. Using density functional theory in conjunction with Wannier interpolation techniques, we study the phonon properties and electron phonon interactions in La$_3$Ni$_2$O$_7$ under 29.5 GPa. Our findings reveal that the electron phonon coupling is insufficient to solely explain the observed high superconducting $T_c$ $\sim$ 80 K in La$_3$Ni$_2$O$_7$. And the calculated strong Fermi surface nesting may explain the experimental observed charge density wave transition in La$_3$Ni$_2$O$_7$. Our calculations substantiate La$_3$Ni$_2$O$_7$ is an unconventional superconductor.
\end{abstract}


\maketitle

\section{Introduction}
Very recently, superconductivity was observed in Ruddlesden-Popper bilayer perovskite nicklate La$_3$Ni$_2$O$_7$ with a maximum $T_c$ of 80 K above 14 GPa, marking the second occurrence of superconductivity in the nicklate family \cite{Sun-nature621}. In 2019, superconductivity was found in hole-doped infinite-layer RNiO$_2$ thin film (R = La, Nd). With 20\% Sr doping, R$_{0.8}$Sr$_{0.2}$NiO$_2$ shows superconducting $T_c$ in the range of 9-15 K \cite{Li-nature572}. The discovery of superconductivity in La$_3$Ni$_2$O$_7$ raises the superconducting $T_c$ of nickelates to the liquid nitrogen temperature zone, representing a notable advancement in high $T_c$ superconductivity following the discoveries in cuprates and iron-based superconductors.

According to Sun \emph{et al.}~\cite{Sun-nature621}, La$_3$Ni$_2$O$_7$ is a paramagnetic metal. The structure transforms from $Amam$ to $Fmmm$ above 15 GPa. DFT calculations suggest that superconductivity emerges with the metallization of the $\sigma$-bonding states under the Fermi level. Using Wannier downfolding of the density functional theory calculations, Luo \emph{et al.}~\cite{Luo-PRL131} studied a bilayer two-orbital model and captured key ingredients such as band structures and Fermi surfaces for La$_3$Ni$_2$O$_7$. There are two electron-type pockets and one hole-type pocket on the Fermi surfaces. Among them, the two electron-type pockets are composed of Ni-3$d_{x^2-y^2}$ and 3$d_{z^2}$ orbitals, while the hole-type pocket is associated with the Ni-3$d_{z^2}$ orbital. This has been confirmed by subsequent theoretical works where the $e_g$ orbitals of Ni dominate the bands near the Fermi level~\cite{Zhang-PRB108,Ouyang-PRB109,Cao-PRB109,Chen-arXiv,Christiansson-PRL131,Gu-arXiv,Lechermann-PRB108,Sakakibara-PRL132,Shilenko-PRB108,Yang-PRB108}. Although long-range magnetic order has not been observed in La$_3$Ni$_2$O$_7$, a large interorbital hopping between 3$d_{z^2}$ and 3$d_{x^2-y^2}$ via in-plane O-2$p_{x/y}$ orbitals is found, which suggests a possible in-plane ferromagnetic tendency~\cite{Zhang-PRB108}. Additionally, the electronic correlation has been extensively investigated. The DFT+DMFT calculations show a remarkable orbital-selective renormalization of Ni 3$d$ orbitals~\cite{Shilenko-PRB108}. Correlated electronic structure and concomitant superconductivity instability originate from the interplay of half-filled Ni-$e_g$ orbitals~\cite{Lechermann-PRB108}. Both spin-orbital separation and spin-frozen phase have also been found in DFT+DMFT calculations, which confirms the nature of Hund metal for La$_3$Ni$_2$O$_7$~\cite{Ouyang-PRB109}. The Hund's rule coupling enhances the quasiparticle effective mass, and their lifetimes are inversely proportional to the temperature, explaining the strange metal behavior experimentally observed in the normal state~\cite{Cao-PRB109}.
Concerning superconductivity, a possible $s_{\pm}$ wave superconductivity pairing in La$_3$Ni$_2$O$_7$ has been proposed~\cite{Yang-PRB108}. And both model studies by functional renormalization group method~\cite{Gu-arXiv} and cluster dynamical mean-field theory~\cite{Tian-arXiv} show the $s_{\pm}$-wave pairing tendency.
Besides, based on local inter-layer spin singlets of Ni-3$d_{z^2}$ orbitals, a minimal effective model containing local pairing of 3$d_{z^2}$ electrons and considerable hybridization with near quarter-filled itinerant 3$d_{x^2-y^2}$ electrons has been proposed, highlighting the importance of the bilayer structure of superconducting La$_3$Ni$_2$O$_7$~\cite{Yang2-PRB108}.

It can be seen that the Ni-3$d_{z^2}$ orbital plays a crucial role in the superconductivity of La$_3$Ni$_2$O$_7$. Although electronic correlation and unconventional superconducting pairing mechanisms have been extensively studied, the potential electron-phonon coupling superconductivity brought by the metallic $\sigma$ bands formed by the Ni-3$d_{z^2}$ and O-2$p_z$ orbitals should not be ignored. It is well-known that the metallization of $\sigma$ bands transforms MgB$_2$ into a high superconductor with $T_c$ of 39 K~\cite{Nagamatsu-nature410}. In unconventional irons-based superconductors, electron phonon coupling may also play an important role~\cite{Lee-nature515}. Hence, it is natural to inquire whether the metallic $\sigma$ bands cause a strong electron-phonon coupling in La$_3$Ni$_2$O$_7$. If so, how significantly do they contribute to the superconductivity. These aspects require systematic clarification.

In this letter, based on first-principles electronic structure calculations, we systematically investigate the phonon properties and the electron-phonon coupling in La$_3$Ni$_2$O$_7$ under pressure. Our density functional perturbation theory calculations reveal no imaginary phonons, indicating La$_3$Ni$_2$O$_7$ is dynamically stable. The electron phonon coupling parameter $\lambda$ is computed to be 0.14. This suggests that the electron phonon coupling in La$_3$Ni$_2$O$_7$ is too weak to drive the BCS pairing. Although the superconducting $T_c$ is estimated to be 0 K based on the McMillian-Allen-Dynes formula, we find strong Fermi surface nesting in La$_3$Ni$_2$O$_7$. This may explain that the charge density wave (CDW) observed in the experimental is due to the Fermi surface nesting.

\begin{figure}[bht]
\centering
\includegraphics[width=8.6cm]{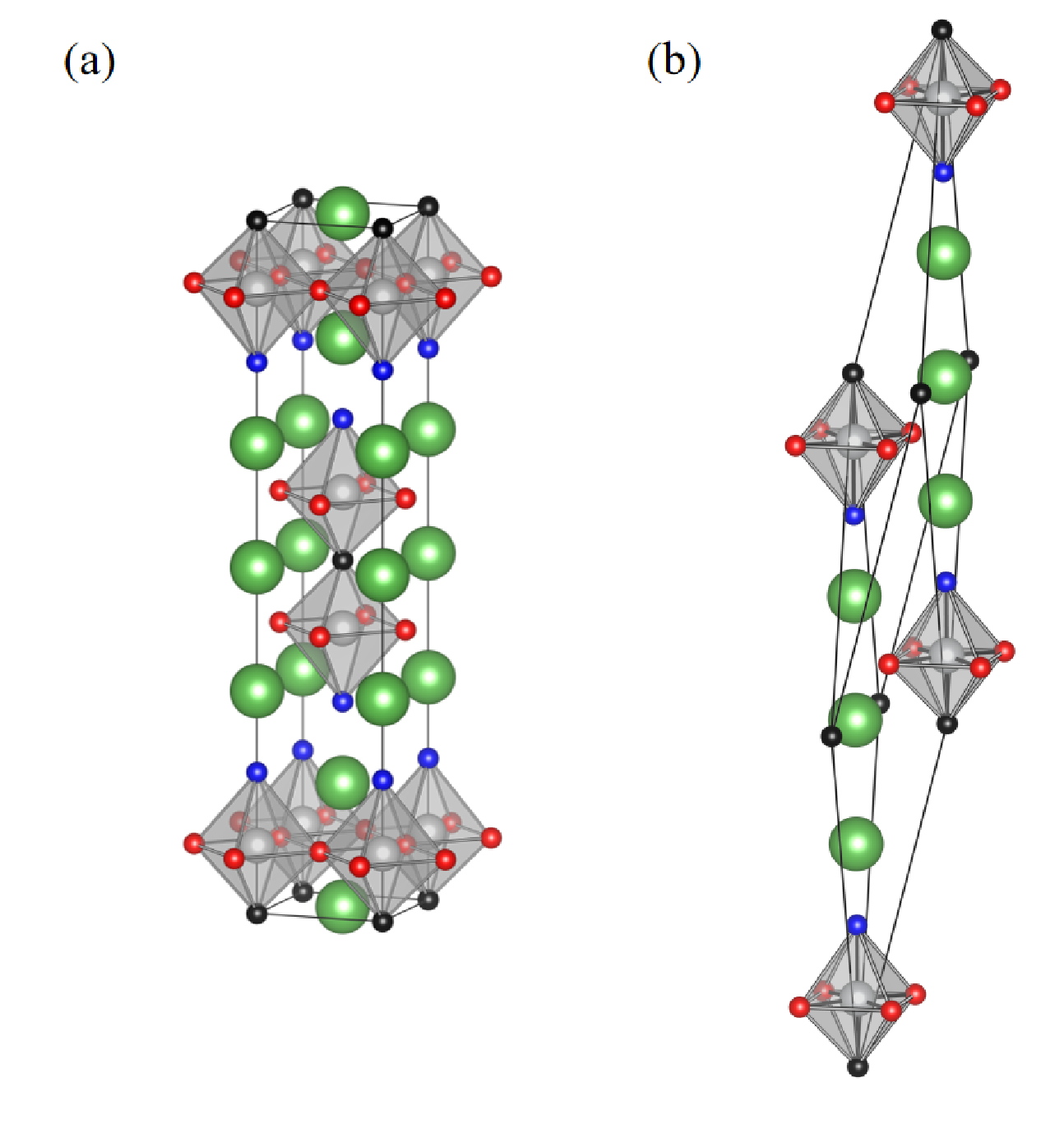}
\caption{The crystal structure for $I4/mmm$ La$_3$Ni$_2$O$_7$, (a) crystal cell and (b) unit cell. The green, gray, black, red and blue balls denote La, Ni, the apical O atoms linking two Ni-O octahedrons, the in-plane O atoms and the apical O atoms locating at the surface of layer, respectively. }
\label{fig:structure}
\end{figure}

\section{Method}
In our calculations, the plane-wave method, QUANTUM-ESPRESSO~\cite{Giannozzi-JPCM21} package was adopted. We calculated the electronic structure and phonon perturbation potentials using the generalized gradient approximation (GGA) of Perdew-Burke-Ernzerhof formula~\cite{Perdew-PRL77} and the optimized norm-conserving Vanderbilt pseudopotentials~\cite{Hemann-PRB88}. The kinetic energy and the charge densities cutoff were chosen to be 80 Ry and 320 Ry, respectively. The charge density were calculated on an unshifted $\bf k$ mesh of 12$\times$12$\times$12 points in combination with a Methfessel-Paxton smearing~\cite{Methfessel-PRB40} of 0.02 Ry. The dynamical matrices and the perturbation potentials were computed on a $\Gamma$-centered 4$\times$4$\times$4 mesh, within the framework of density functional perturbation theory~\cite{Baroni-RMP73}.

We constructed the maximally localized Wannier functions (MLWFs)~\cite{Pizzi-JPCM32} on a 4$\times$4$\times$4 grid of the Brillouin zone. Three $p$ orbitals of O atoms and five $d$ orbitals were included during the generation of MLWFs for La$_3$Ni$_2$
O$_7$. After optimization, these MLWFs exhibit excellent localization in space. For example, the average spatial spread of MLWFs for La$_3$Ni$_2$ is 1.05 $\AA^2$. The convergence test for electron-phonon coupling (EPC) constatnt $\lambda$ was carried out through fine electron (32$\times$32$\times$32) and phonon (16$\times$16$\times$16) grids with EPW codes~\cite{Ponce-CPC209}. The Dirac $\delta$ functions for electrons and phonons were smeared out by a Gaussian function with widths of 90 meV and 0.5 meV, respectively.

The mode- and wavevecter-dependent coupling $\lambda_{\textbf{q}\nu}$ reads
\begin{equation}
\lambda_{\textbf{q}\nu} = \frac{2}{{\hbar}N(0)N_{\textbf{k}}} \sum_{nm{\textbf{k}}} \frac{1}{{\omega}_{{\textbf{q}\nu}}} {\vert {g^{nm}_{{\textbf{k},{\textbf{q}}}\nu}} \vert}^2 \delta(\epsilon^{n}_{\textbf{k}}) \delta(\epsilon^{m}_{\textbf{k+q}}),
\end{equation}
with which the strongly coupled phonon modes can be identified. $N$(0) is the density of states (DOS) of electrons at the Fermi level. $N_{\textbf{k}}$ is the total number of {\textbf{k}} points in  the fine {\textbf{k}}-mesh. $\omega_{{\textbf{q}}\nu}$ is the phonon frequency and $g^{nm}_{{\textbf{k},{\textbf{q}}}\nu}$ is the EPC matrix element. ($n$,$m$) and $\nu$ denote the indices of energy bands and phonon mode, respectively. $\epsilon^{n}_{\textbf{k}}$ and $\epsilon^{m}_{\textbf{k+q}}$ are the eigenvalues of the Kohn-Sham orbitals with respect to the Fermi level.
\par
The EPC constant $\lambda$ was determined by the summation of $\lambda_{{\textbf{q}}\nu}$ over the first Brillouin zone, or the intergration of the Eliahberg spectral function $\alpha^2F(\omega)$,
\begin{equation}
\lambda = \frac{1}{N_{\textbf{q}}} \sum_{{\textbf{q}}\nu} \lambda_{\textbf{q}\nu} = 2 \int{\frac{\alpha^2F(\omega)}{\omega}} d\omega,
\end{equation}
where $N_\textbf{q}$ represents the total number of \textbf{q} points in the fine \textbf{q}-mesh. The Eliashberg spectral function $\alpha^2F(\omega)$ was calaulated with
\begin{equation}
\alpha^2F(\omega) = \frac{1}{2N_{\textbf{q}}} \sum_{\textbf{q}\nu} \lambda_{\textbf{q}\nu}\omega_{\textbf{q}\nu}\delta(\omega - \omega_{{\textbf{q}}\nu}).
\end{equation}
The superconducting transition temperature is determined by the McMillian-Allen-Dynes fomula~\cite{Allen-PRB6,Allen-PRB12},
\begin{equation}
T_c = \frac{\omega_{log}}{1.2}\exp\left[{\frac{-1.04(1+\lambda)}{\lambda(1 - 0.62\mu^*) - \mu^*}}\right]
\end{equation}
in which $\mu^*$  is the Coulomb pseudopotential, and $\omega_{\text{log}}$ is the logarithmic average frequency that is defined as
\begin{equation}
\omega_{\text{log}} = \exp\left[\frac{2}{\lambda}\int{\frac{d\omega}{\omega}\alpha^2F(\omega)ln(\omega)}\right],
\end{equation}
The Fermi surface nesting function is calculated with
\begin{equation}
\xi(\mathbf{q})=\frac{1}{N(0)N_\mathbf{k}}\sum_{nm\mathbf{k}}\delta(\epsilon _{\mathbf{k}}^{n})\delta (\epsilon _{\mathbf{k+q}}^{m}).
\end{equation}

\begin{figure*}[tbh]
\centering
\includegraphics[width=17.2cm]{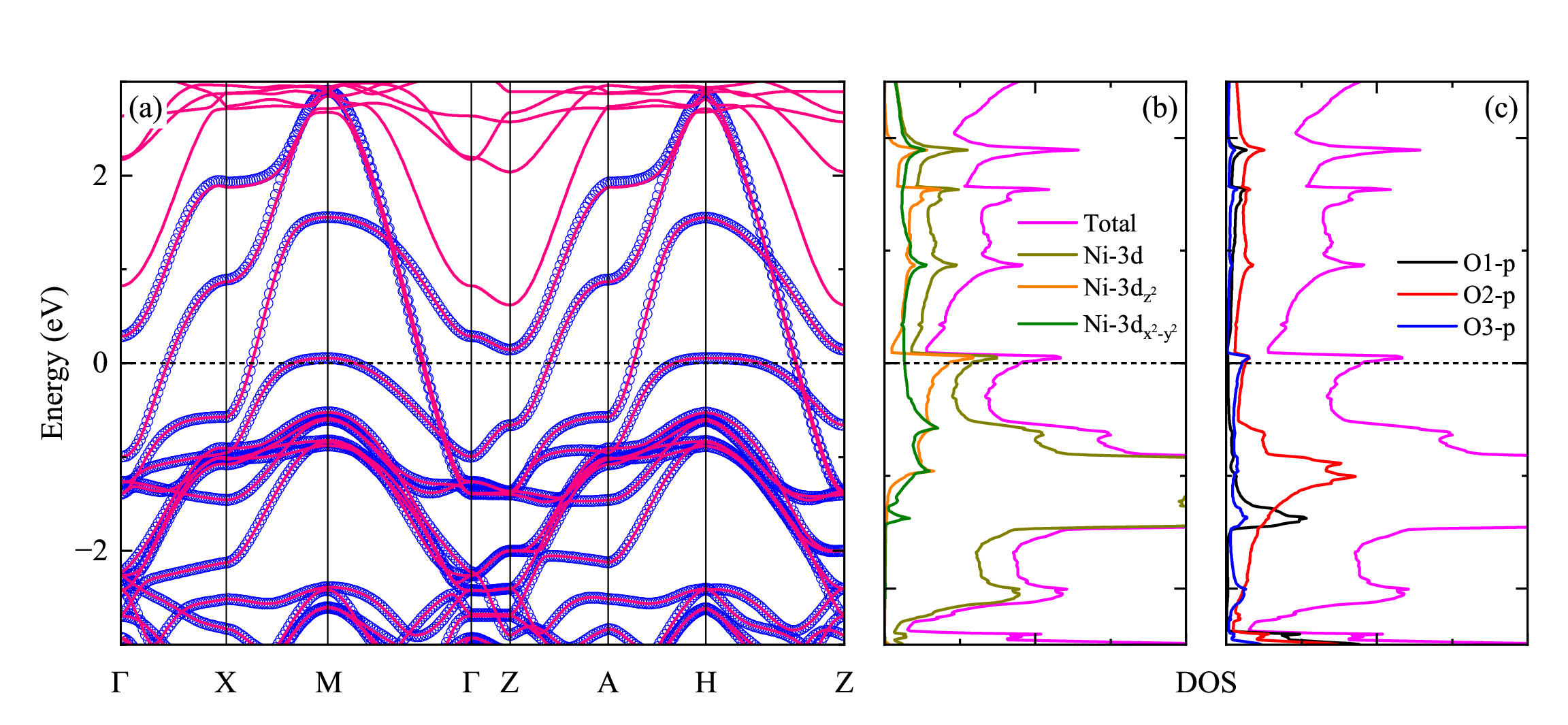}
\includegraphics[width=17.2cm]{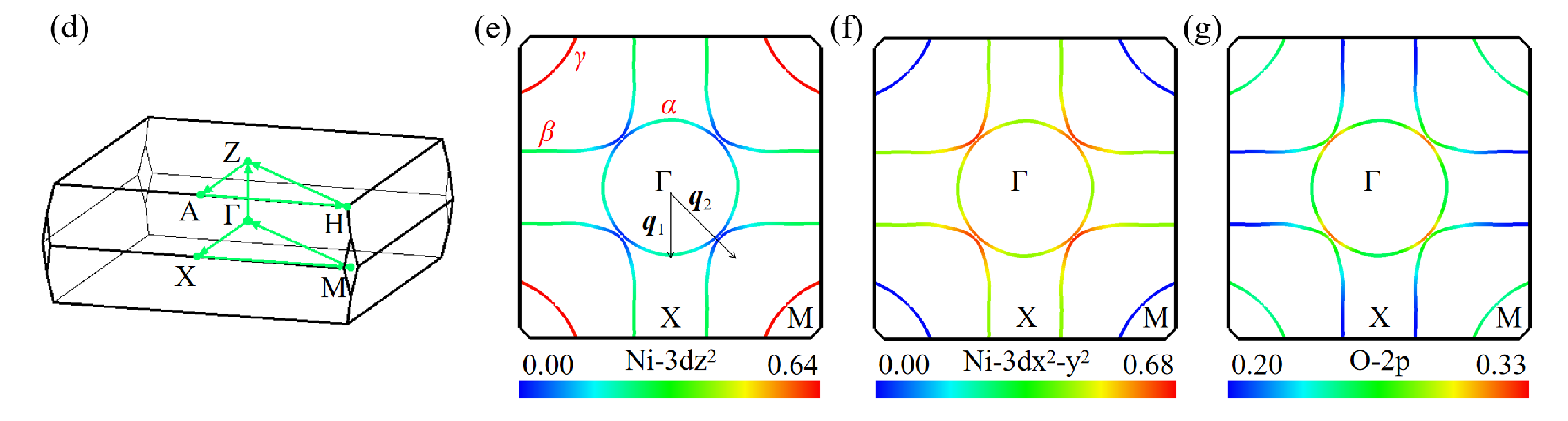}
\caption{Electronic structure of La$_3$Ni$_2$O$_7$. (a) Bands structure. The Fermi level is set to zero. The pink lines and blue circles represent the DFT energy bands and Wannier energy bands, respectively. (b) and (c) Total and orbital-resolved DOS. (d) The Brillouin zone and the high symmetry path. (e)-(g) Distributions of different orbitals on the $k_z$ = 0 Fermi surfaces. }
\label{fig:band}
\end{figure*}

\begin{figure*}[t]
\centering
\includegraphics[width=17.2cm]{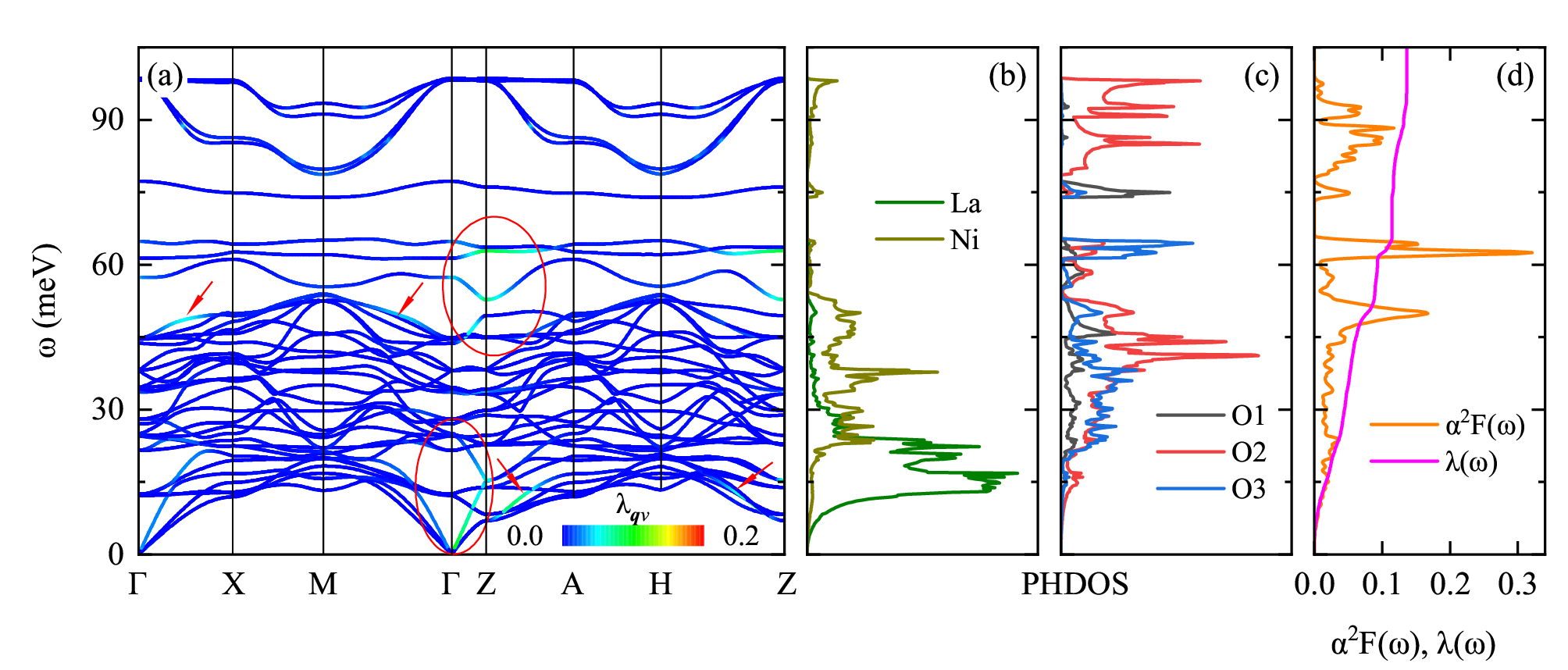}
\caption{(a) Phonon spectrum with a color representation of $\lambda_{qv}$ for La$_3$Ni$_2$O$_7$. (b) and (c) Projected phonon DOS generated by quasiharmonic approximation. (d) Eliashberg spectral function $\alpha^2F(\omega)$ and accumulated EPC constant $\lambda(\omega)$.}
\label{fig:phonon}
\end{figure*}

\begin{figure}[b]
\includegraphics[width=8.6cm]{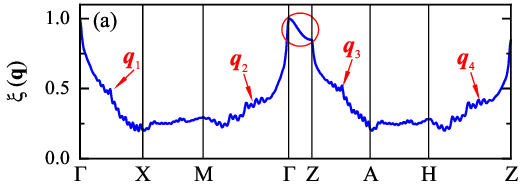}
\includegraphics[width=8.6cm]{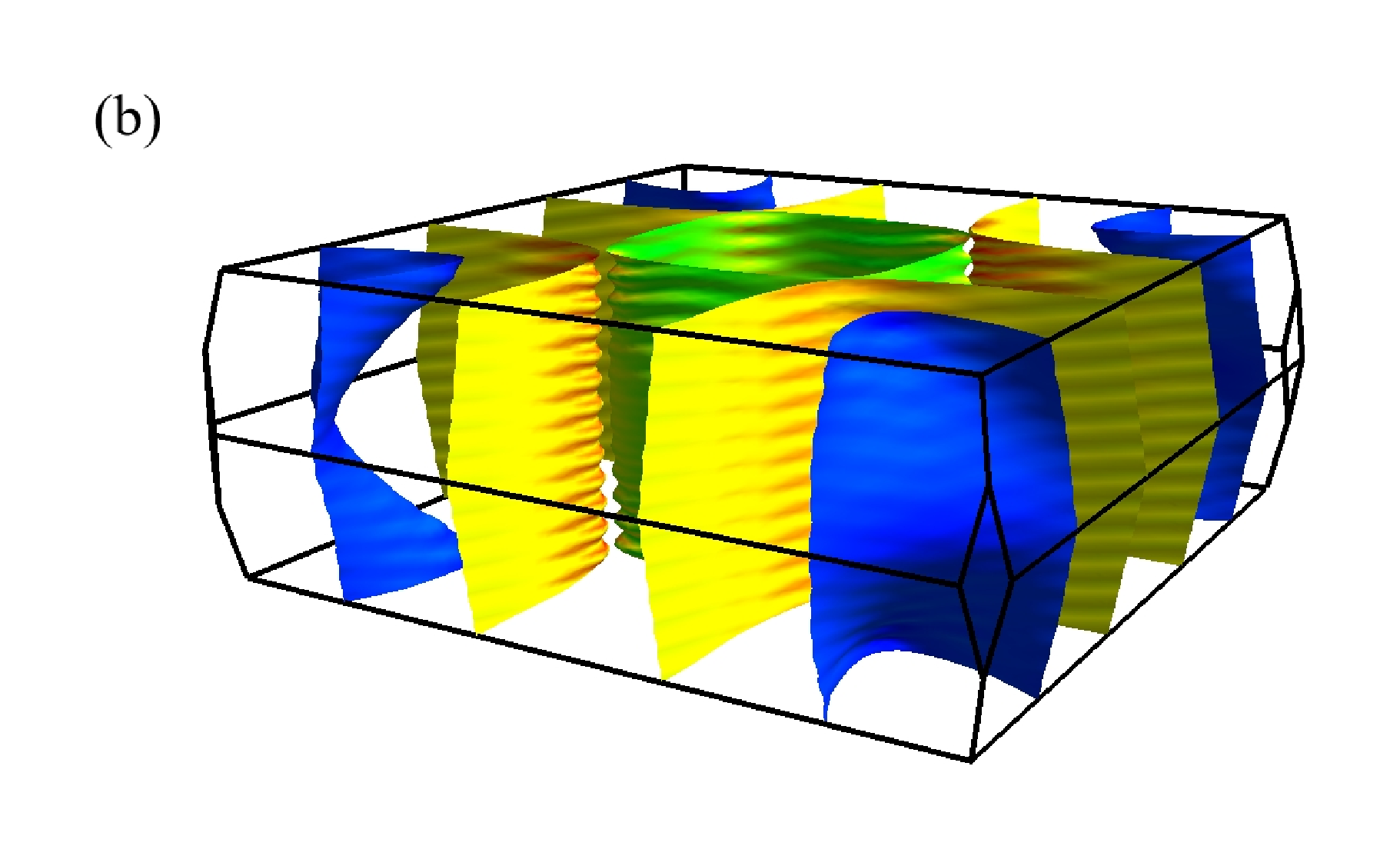}
\centering
\caption{(a) The Fermi surface nesting function $\xi(\mathbf{q})$ along the high symmetry path, which is normalized by $\xi(\Gamma)$. (b) The three-dimensional Fermi surfaces in the first Brillouin zone }
\label{fig:nesting}
\end{figure}
\section{Results and Discussions}

According to Sun \emph{et al.}~\cite{Sun-nature621}, the space group for La$_3$Ni$_2$O$_7$ is $Fmmm$ under 29.5 GPa, where the lattice constants $a_{\text{exp}}$ and $b_{\text{exp}}$ are 5.289 and 5.218 \AA, respectively. However, our optimized results indicate that $a_{\text{opt}}$ and $b_{\text{opt}}$ for the $Fmmm$ structure are almost equal, approximately 5.20579 \AA. We initially chose a body-centered tetragonal $I4/mmm$ structure ($a^* = b^*$) for our calculations. Upon full relaxation, the lattice constant $a^*$ for $I4/mmm$ structure is 3.680 $\AA$, representing the distance between in-plane La atoms. This value closely aligns with the optimized $Fmmm$ value of 3.681 \AA, as well as the experimental value of 3.715 $\AA$. The maximum error does not exceed 1\%. Hence, we adopted the body-centered tetragonal $I4/mmm$ structure La$_3$Ni$_2$O$_7$ in our calculations. The crystal structure and the unit cell for $I4/mmm$  La$_3$Ni$_2$O$_7$ are shown in Fig.~\ref{fig:structure}. The Ni-O octahedron includes three different O atoms. For convenience, we label the apical O atoms linking two Ni-O octahedrons as O1, the in-plane O atoms as O2 and the apical O atoms locating at the surface of layer as O3, which are represented by black, red, and blue balls in Fig.~\ref{fig:structure}, respectively.

We then study the electronic structure of La$_3$Ni$_2$O$_7$ under 29.5 GPa. Figure.~\ref{fig:band} shows the band structure, density of states, and the Fermi surfaces. We also construct the tight-binding model. The wannier bands are in great consistence with our DFT bands, as well as the previous theoretic work~\cite{Luo-PRL131}. We see three bands acrossing the Fermi level. Among them, the bonding states and the anti-bonding states formed by the Ni-3$d_{z^2}$ and O-2$p_z$ orbitals are splitted at the $M$ and $H$ points. The anti-bonding states locate at about 1.5 eV above the Fermi level, while the bonding states is corresponding to the flat band at the Fermi level. Such a flat band contributes a significant DOS peak at the Fermi level [Fig.~\ref{fig:band}(b) and (c)]. The atomic orbital-projected DOS suggests that the Ni-3$d_{z^2}$ orbitals dominate the DOS peak around the Fermi level. In Fig.~\ref{fig:band}(e)-(g), we exhibit the orbital-resolved Fermi surfaces at $k_{z} = 0$ plane. We see three Fermi pockets, which are corresponding to those three bands acrossing the Fermi level. The electron-type Fermi surfaces $\alpha$ and $\beta$ around the $\Gamma$ and $X$ points are contributed by both the 3$d_{z^2}$ and 3$d_{x^2-y^2}$ orbitals. And the hole-type $\gamma$ pocket around the $M$ point is mainly composed of the Ni-3$d_{z^2}$ and O-2$p$ orbitals. All these findings indicate that there are metalized $\sigma$ bands in La$_3$Ni$_2$O$_7$. It is necessary to check what role the electron phonon coupling plays in superconductor La$_3$Ni$_2$O$_7$ with $T_c$ $\sim$ 80 K under pressure.

We next investigate the phonon properties and the electron-phonon coupling of La$_3$Ni$_2$O$_7$. Figure.~\ref{fig:phonon}(a)-(c) show the phonon spectrum and the projected phonon DOS. No imaginary phonon mode is observed. This suggests that La$_3$Ni$_2$O$_7$ is dynamically stable under 29.5 GPa. We learn that the lowest acoustic phonons are mainly composed by La atoms. And the medium frequency branches from 22 to 60 meV are mixed by Ni, O atoms. We also see a gap around 75 meV. The O1 and O2 related phonons dominate the high frequency zone above the gap. The calculated Eliashberg spectral function $\alpha^2F(\omega)$ and integrated EPC constant $\lambda$ for La$_3$Ni$_2$O$_7$ are shown in Fig.~\ref{fig:phonon}(d). Two obvious $\alpha^2F(\omega)$ peaks are found around 50 and 60 meV. However, the integrated EPC $\lambda$ is computed to be 0.14. Using the McMillan-Allen-Dynes formula, we obtain the superconducting transition $T_c$ for La$_3$Ni$_2$O$_7$ is to be 0 K, when the Coulomb pseudopotential $\mu^*$ is set as 0.1. This finding indicates that the EPC in La$_3$Ni$_2$O$_7$ under pressure is weak. The phonon spectrum with $\lambda_{\mathbf{q}v}$ in Fig.~\ref{fig:phonon}(a) also confirms this fact. At first glance, the EPC $\lambda_{\mathbf{q}v}$ of almost all phonons are negligible. And only at some special $\mathbf{q}$ points, $\lambda_{\mathbf{q}v}$ have finite intensities, which are labeled by the red arrows or circles in Fig.~\ref{fig:phonon}(a). It's very reminiscent of Fermi surface nesting or phonon softening that may enhance the EPC $\lambda_{\mathbf{q}v}$.

Thus, we further study the Fermi surface nesting function $\xi(\mathbf{q})$ for La$_3$Ni$_2$O$_7$. As shown in Fig.~\ref{fig:nesting}(a), the red arrows and red circle all mark the $\mathbf{q}$ vectors with strong Fermi surface nesting. The previously labelled $\mathbf{q}$ points of finite EPC $\lambda_{\mathbf{q}v}$ coincide perfectly with these $\mathbf{q}$ vectors. Among them, a high peak of $\xi(\mathbf{q})$ is found along the $\Gamma$-$Z$ path, which suggests that La$_3$Ni$_2$O$_7$ has a two-dimensional characteristic. The three-dimensional display of the Fermi surfaces in Fig.~\ref{fig:nesting}(b) also confirms this fact. All the Fermi surfaces keep good nesting along the $c$ axis, except the hourglass-shaped $\gamma$ pockets. In addition, there are also strong in-plane Fermi surface nesting in La$_3$Ni$_2$O$_7$. For example, the in-plane $\mathbf{q}$ vectors $\mathbf{q}_1$, $\mathbf{q}_2$, $\mathbf{q}_3$, and $\mathbf{q}_4$ possess high intensities of $\xi(\mathbf{q})$ and contribute a finite EPC $\lambda_{\mathbf{q}v}$ as marked in Fig.~\ref{fig:phonon}(a). Due to the quasi-two-dimensional Fermi surfaces, $\mathbf{q}_1$ ($\mathbf{q}_2$) vector shares the same in-plane component with $\mathbf{q}_3$ ($\mathbf{q}_4$). And the vectors $\mathbf{q}_1$ and $\mathbf{q}_2$ are exhibited in Fig.~\ref{fig:band}(e). The strong Fermi surface nesting may explain the experimental observed CDW transition in La$_3$Ni$_2$O$_7$~\cite{Liu-SC66}, since the EPC strength is negligible according to our calculations.

\section{CONCLUSION}
In summary, we have performed the first-principles density functional theory calculations to study the electronic structure of La$_3$Ni$_2$O$_7$ under 29.5 GPa. As shown by our calculations,  La$_3$Ni$_2$O$_7$ is a metal and the Fermi surfaces are mainly constructed by the 3$d_{z^2}$ and 3$d_{x^2-y^2}$ orbitals of Ni. We also have investigated the lattice dynamics and checked the electron phonon coupling in La$_3$Ni$_2$O$_7$. Our results suggest that La$_3$Ni$_2$O$_7$ is an unconventional high-$T_c$ superconductor. The superconductivity with $T_c$ $\sim$ 80 K in La$_3$Ni$_2$O$_7$ goes beyond the frame of electron phonon coupling, which may be totally attributed to the strong electronic correlation. And the calculated strong Fermi surface nesting may explain the experimental observed CDW transition in La$_3$Ni$_2$O$_7$. Our work qualitatively excludes the possibility of electron phonon coupling mechanism in the superconducting phase of La$_3$Ni$_2$O$_7$ under pressure. Moreover, considering the absence of electron phonon mediated superconductivity in infinite-layer nickelates~\cite{Cataldo-PRB108, CPL-Hou39}, our results once again confirm that superconductivity in nickelates is possible via unconventional Cooper pairing.

~\newline

$\emph{Note added}$: During the preparation of this manuscript, Ref.~\cite{Talantsev-arXiv, Li-arXiv, Yi-arXiv} appear on the arXiv preprint server studying the electron phonon coupling (EPC) in La$_3$Ni$_2$O$_7$.

\begin{acknowledgments}

This work was supported by the National Natural Science Foundation of China (Grant No. 11934020).
Computational resources were provided by the Physical Laboratory of High Performance Computing at Renmin University of China.

\end{acknowledgments}

\end{document}